\documentclass[twoside]{dis09}
\usepackage[latin1]{inputenc}
\usepackage[dvips]{graphicx,epsfig,color}
\usepackage{wrapfig,rotating}
\usepackage{amssymb,amsmath,array}

\begin{document}

\newcommand{\pom}{{I\!\!P}}
\newcommand{\reg}{{I\!\!R}}
\newcommand{\slowpi}{\pi_{\mathit{slow}}}
\newcommand{\fiidiii}{F_2^{D(3)}}
\newcommand{\fiidiiiarg}{\fiidiii\,(\beta,\,Q^2,\,x)}
\newcommand{\n}{1.19\pm 0.06 (stat.) \pm0.07 (syst.)}
\newcommand{\nz}{1.30\pm 0.08 (stat.)^{+0.08}_{-0.14} (syst.)}
\newcommand{\fiidiiiful}{F_2^{D(4)}\,(\beta,\,Q^2,\,x,\,t)}
\newcommand{\fiipom}{\tilde F_2^D}
\newcommand{\ALPHA}{1.10\pm0.03 (stat.) \pm0.04 (syst.)}
\newcommand{\ALPHAZ}{1.15\pm0.04 (stat.)^{+0.04}_{-0.07} (syst.)}
\newcommand{\fiipomarg}{\fiipom\,(\beta,\,Q^2)}
\newcommand{\pomflux}{f_{\pom / p}}
\newcommand{\nxpom}{1.19\pm 0.06 (stat.) \pm0.07 (syst.)}
\newcommand {\gapprox}
   {\raisebox{-0.7ex}{$\stackrel {\textstyle>}{\sim}$}}
\newcommand {\lapprox}
   {\raisebox{-0.7ex}{$\stackrel {\textstyle<}{\sim}$}}
\def\gsim{\,\lower.25ex\hbox{$\scriptstyle\sim$}\kern-1.30ex%
\raise 0.55ex\hbox{$\scriptstyle >$}\,}
\def\lsim{\,\lower.25ex\hbox{$\scriptstyle\sim$}\kern-1.30ex%
\raise 0.55ex\hbox{$\scriptstyle <$}\,}
\newcommand{\pomfluxarg}{f_{\pom / p}\,(x_\pom)}
\newcommand{\dsf}{\mbox{$F_2^{D(3)}$}}
\newcommand{\dsfva}{\mbox{$F_2^{D(3)}(\beta,Q^2,x_{I\!\!P})$}}
\newcommand{\dsfvb}{\mbox{$F_2^{D(3)}(\beta,Q^2,x)$}}
\newcommand{\dsfpom}{$F_2^{I\!\!P}$}
\newcommand{\gap}{\stackrel{>}{\sim}}
\newcommand{\lap}{\stackrel{<}{\sim}}
\newcommand{\fem}{$F_2^{em}$}
\newcommand{\tsnmp}{$\tilde{\sigma}_{NC}(e^{\mp})$}
\newcommand{\tsnm}{$\tilde{\sigma}_{NC}(e^-)$}
\newcommand{\tsnp}{$\tilde{\sigma}_{NC}(e^+)$}
\newcommand{\st}{$\star$}
\newcommand{\sst}{$\star \star$}
\newcommand{\ssst}{$\star \star \star$}
\newcommand{\sssst}{$\star \star \star \star$}
\newcommand{\tw}{\theta_W}
\newcommand{\sw}{\sin{\theta_W}}
\newcommand{\cw}{\cos{\theta_W}}
\newcommand{\sww}{\sin^2{\theta_W}}
\newcommand{\cww}{\cos^2{\theta_W}}
\newcommand{\trm}{m_{\perp}}
\newcommand{\trp}{p_{\perp}}
\newcommand{\trmm}{m_{\perp}^2}
\newcommand{\trpp}{p_{\perp}^2}
\newcommand{\alp}{\alpha_s}

\newcommand{\alps}{\alpha_s}
\newcommand{\sqrts}{$\sqrt{s}$}
\newcommand{\LO}{$O(\alpha_s^0)$}
\newcommand{\Oa}{$O(\alpha_s)$}
\newcommand{\Oaa}{$O(\alpha_s^2)$}
\newcommand{\PT}{p_{\perp}}
\newcommand{\JPSI}{J/\psi}
\newcommand{\sh}{\hat{s}}
\newcommand{\uh}{\hat{u}}
\newcommand{\MP}{m_{J/\psi}}
\newcommand{\PO}{I\!\!P}
\newcommand{\xbj}{x}
\newcommand{\xpom}{x_{\PO}}
\newcommand{\ttbs}{\char'134}
\newcommand{\xpomlo}{3\times10^{-4}}  
\newcommand{\xpomup}{0.05}  
\newcommand{\dgr}{^\circ}
\newcommand{\pbarnt}{\,\mbox{{\rm pb$^{-1}$}}}
\newcommand{\gev}{\,\mbox{GeV}}
\newcommand{\WBoson}{\mbox{$W$}}
\newcommand{\fbarn}{\,\mbox{{\rm fb}}}
\newcommand{\fbarnt}{\,\mbox{{\rm fb$^{-1}$}}}
\newcommand{\dsdx}[1]{$d\sigma\!/\!d #1\,$}
\newcommand{\eV}{\mbox{e\hspace{-0.08em}V}}
%
%
\newcommand{\qsq}{\ensuremath{Q^2} }
\newcommand{\gevsq}{\ensuremath{\mathrm{GeV}^2} }
\newcommand{\et}{\ensuremath{E_t^*} }
\newcommand{\rap}{\ensuremath{\eta^*} }
\newcommand{\gp}{\ensuremath{\gamma^*}p }
\newcommand{\dsiget}{\ensuremath{{\rm d}\sigma_{ep}/{\rm d}E_t^*} }
\newcommand{\dsigrap}{\ensuremath{{\rm d}\sigma_{ep}/{\rm d}\eta^*} }

\newcommand{\dstar}{\ensuremath{D^*}}
\newcommand{\dstarp}{\ensuremath{D^{*+}}}
\newcommand{\dstarm}{\ensuremath{D^{*-}}}
\newcommand{\dstarpm}{\ensuremath{D^{*\pm}}}
\newcommand{\zDs}{\ensuremath{z(\dstar )}}
\newcommand{\Wgp}{\ensuremath{W_{\gamma p}}}
\newcommand{\ptds}{\ensuremath{p_t(\dstar )}}
\newcommand{\etads}{\ensuremath{\eta(\dstar )}}
\newcommand{\ptj}{\ensuremath{p_t(\mbox{jet})}}
\newcommand{\ptjn}[1]{\ensuremath{p_t(\mbox{jet$_{#1}$})}}
\newcommand{\etaj}{\ensuremath{\eta(\mbox{jet})}}
\newcommand{\detadsj}{\ensuremath{\eta(\dstar )\, \mbox{-}\, \etaj}}

\def\Journal#1#2#3#4{{#1} {\bf #2} (#3) #4}
\def\NCA{\em Nuovo Cimento}
\def\NIM{\em Nucl. Instrum. Methods}
\def\NIMA{{\em Nucl. Instrum. Methods} {\bf A}}
\def\NPB{{\em Nucl. Phys.}   {\bf B}}
\def\PLB{{\em Phys. Lett.}   {\bf B}}
\def\PRL{\em Phys. Rev. Lett.}
\def\PRD{{\em Phys. Rev.}    {\bf D}}
\def\ZPC{{\em Z. Phys.}      {\bf C}}
\def\EJC{{\em Eur. Phys. J.} {\bf C}}
\def\CPC{\em Comp. Phys. Commun.}


\newcommand{\m}{\ensuremath{\mathrm{m}}}
\newcommand{\cm}{\ensuremath{\mathrm{cm}}}
\newcommand{\um}{\ensuremath{\mathrm{\mu m}}}
\newcommand{\us}{\ensuremath{\mathrm{\mu s}}}
\newcommand{\pb}{\ensuremath{\mathrm{pb}}}
\newcommand{\GeV}{\ensuremath{\mathrm{GeV}}}
\newcommand{\MeV}{\ensuremath{\mathrm{MeV}}}
\newcommand{\der}{\ensuremath{\mathrm{d}}}
\newcommand{\ns}{\ensuremath{\mathrm{ns}}}
\newcommand{\xxx}{\ensuremath{\mathrm{xxx}}}
\newcommand{\rem}[1]{{\bfseries \itshape #1}}
\newcommand{\sub}[1]{\ensuremath{_\mathrm{#1}}}
\newcommand{\super}[1]{\ensuremath{^\mathrm{#1}}}
\newcommand{\pt}{p\sub T}
\newcommand{\ptrel}{p\sub T \super{rel}}
\newcommand{\pz}{p\sub z}
\newcommand{\ptmu}{p\sub T ^\mu}
\newcommand{\ptjet}{p\sub T \super{jet}}
\newcommand{\ptjets}{p\sub T \super{jet 1 (2)}}
\newcommand{\ptjetone}{p\sub T \super{jet 1}}
\newcommand{\ptjettwo}{p\sub T \super{jet 2}}
\newcommand{\etamu}{\eta^\mu}
\newcommand{\etajet}{\eta\super{jet}}
\newcommand{\etajets}{\eta\super{jet 1 (2)}}
\newcommand{\dEdx}{\der E / \der x}
\newcommand{\mK}{m\sub {K}}
\newcommand{\mKK}{m\sub {KK}}
\newcommand{\mphi}{m\sub {\phi}}
\newcommand{\nS}{n\sub {S}}
\newcommand{\Nbin}{N\sub {bin}}
\newcommand{\Nb}{N\sub {b}}
\newcommand{\fb}{f\sub {b}}
\newcommand{\effrec}{\epsilon\sub {rec}}
\newcommand{\efftrig}{\epsilon\sub {trig}}
\newcommand{\sigmaep}{\sigma\sub{ep}}
\newcommand{\sigmagammap}{\sigma\sub{\gamma p}}
\newcommand{\sigmagammapT}{\sigma\sub{\gamma p}\super{T}}
\newcommand{\sigmagammapL}{\sigma\sub{\gamma p}\super{L}}
\newcommand{\phiL}{\phi\sub{\gamma/e}\super{L}}
\newcommand{\phiT}{\phi\sub{\gamma/e}\super{T}}
\newcommand{\Phigammae}{\Phi\sub{\gamma/e}}
\newcommand{\Lumi}{\ensuremath{\mathcal{L}}}
\newcommand{\Lint}{\ensuremath{\int {\mathcal{L} \der t}}}
\newcommand{\MY}{\ensuremath{M\sub{Y}}}
\newcommand{\Ntag}{\ensuremath{N\sub{tag}}}
\newcommand{\Nuntag}{\ensuremath{N\sub{untag}}}
\newcommand{\ftag}{\ensuremath{f\sub{tag}}}
\newcommand{\epsel}{\ensuremath{\epsilon\sub{el}}}
\newcommand{\epspd}{\ensuremath{\epsilon\sub{pd}}}
\newcommand{\Nel}{\ensuremath{N\sub{el}}}
\newcommand{\Npd}{\ensuremath{N\sub{pd}}}
\newcommand{\artitle}[1]{}
\newcommand{\ysigma}{\ensuremath{y \sub \Sigma}}
\newcommand{\xgamma}{\ensuremath{x \sub \gamma \super{obs}}}
\newcommand{\deltaphi}{\ensuremath{\delta\phi\sub{jets}}}
\newcommand{\mur}{\ensuremath{\mu\sub{r}}}
\newcommand{\muf}{\ensuremath{\mu\sub{f}}}

\title{A Measurement of Beauty Photoproduction
Through Decays to Muons and Jets at HERA-II }

\author{Benno List%
%
\thanks{Supported by the German Federal Ministry of Science and
Technology under grant 05H16GUA.} ~for the H1 Collaboration
%
\vspace{.3cm}\\
%
University of Hamburg - Institute for Experimental Physics \\
Luruper Chaussee 149, D--22603 Hamburg - Germany
}

\maketitle

\begin{abstract}
The photoproduction of beauty quarks in ep collisions 
has been measured using a data sample of $170\,\pb^{-1}$ collected
with the H1 detector at HERA-II in the years 2006 and 2007.
Events with
two jets and a muon in the final state were investigated, and
beauty events were identified using the muon's relative transverse momentum to 
a jet and its impact parameter.
Visible cross sections were measured differentially in the transverse
momenta of the highest energy jet ($\ptjetone$) and the muon ($\ptmu$), 
the pseudorapidity of the muon ($\etamu$) and of the photon's momentum fraction
$\xgamma$ entering the hard interaction.
The measurements are found to be well described by QCD calculations at NLO.
\end{abstract}

\section{Introduction}

The production of beauty quarks in $ep$ collisions has been investigated
in considerable detail at HERA-I. In several analyses it was observed that
measured cross sections where significantly above the predictions from 
 perturbative QCD calculations in
next-to-leading order (NLO). The data from HERA-II with its larger
statistics makes it possible to repeat these measurements with increased
accuracy. 

This measurement \cite{url,H1prelim-08-071} follows closely a measurement performed with data from HERA-I
\cite{Aktas:2005zc}, where beauty photoproduction events were investigated with
two jets and a muon in the final state.
The result of the HERA-I measurement was that NLO calculations 
describe the data reasonably well, except
for the lowest bin of the muon and jet transverse momentum, $\ptmu$ and
$\ptjetone$, where the data were significantly above the predictions.
Similar measurement have also been 
made by the ZEUS collaboration \cite{bib:zeus-bmux}, 
in a slightly different phase space. Here, good agreement was found with QCD
predictions, also at low jet and muon $\pt$.

The measurement presented here uses the HERA-II data set to measure the
same cross sections in the same visible range as in the previous H1 
publication with increased
statistics and correspondingly smaller errors.
The data was collected with the H1 detector in the years 2006 and 2007,
when HERA collided electron and positron beams
of an energy of $E\sub e=27.55\,\GeV$ with protons of 
$E\sub p=920\,\GeV$, and
corresponds to an integrated luminosity
of $170\,\pb^{-1}$.

\section{QCD Models}

The Monte Carlo generators PYTHIA 6.2 \cite{bib:pythia} 
and Cascade 2.0 \cite{bib:cascade}
were used for the simulation of signal and background distributions.

The PYTHIA event samples were generated with 
massless matrix elements, using the CTEQ6L \cite{Pumplin:2002vw} and
SAS-1D \cite{Schuler:1996fc}
parton density sets for the proton and the photon,
respectively.
The fragmentation of heavy quarks to hadrons was simulated using the
Peterson fragmentation function
with a parameter $\epsilon\sub{b (c)} = 0.0069 (0.058)$ for beauty
(charm) quarks. More details on the parameter settings can be found in
our previous publication \cite{Aktas:2005zc}.

For cross checks, additional  Monte Carlo samples were generated using
the Cascade program, which  is based on $k\sub T$
factorization and the CCFM evolution rather than collinear factorization
and DGLAP evolution. The proton parton density set A0 \cite{Jung:2004gs} 
is used for the
unintegrated gluon density in the proton. 

Both programs, PYTHIA and Cascade, use Leading Order (LO), i.e.\
$O (\alpha\sub s)$, QCD matrix elements for the hard scattering,
augmented by parton showers that approximate the effect of additional 
multiple gluon emission. 
 
To compare the data to Next-to-Leading Order (NLO) QCD, i.e.\
$O (\alpha\sub s^2)$, the FMNR program \cite{Frixione:1995qc} has been used,
which is based on the NLO calculation by Nason, Dawson, and Ellis
\cite{bib:nde}.
Here, the CTEQ5F4 \cite{Lai:1999wy} (GRV-G HO \cite{Gluck:1991jc}) 
parton density set for the proton
(photon) were employed.
Hadronisation corrections were calculated using the PYTHIA program.

The theory uncertainties were determined by varying the input beauty
mass
$m\sub{b}$ from the nominal value $m\sub{b}=4.75\,\GeV$ up and down by
$0.25\,\GeV$. In addition, the renormalization and factorization scales
$\mur$ and $\muf$ were varied independently in the range
$\mu_0 / 2 \le \mur, \muf \le 2\,\mu_0$, with the constraint 
$1/2 \le \mur/\muf \le 2$, and $\mu_0$ set to 
$\mu_0=\sqrt{p\sub T^2 + m\sub b^2}$.
For each bin on hadron level, the deviations due to the beauty mass 
variation and the largest deviation due to the scale variation in the
upward and downward direction were determined and added in quadrature for the
total model uncertainty, following the prescription in \cite[p. 406]{Alekhin:2005dy}.

\section{Analysis Method}

Events were selected with the following experimental cuts:
No electron candidate with an energy above $E>6\,\GeV$ 
  is allowed to be found in the detector.
The inelasticity $y$, measured from the hadronic activity in the 
  detector, 
  must lie in the range $0.2 < y\sub h < 0.8$.
Two jets with transverse momenta $\ptjets > 7 (6)\,\GeV$
  for the highest (second highest) $\pt$ jet
  have to be found within the pseudorapitity range $-2.5 < \etajet< 2.5$.
  Jets were identified by the inclusive $k\sub t$ jet algorithm
  \cite{bib:ktalgorithm} in the $\pt$ recombination scheme, 
  with a distance parameter 
  $R = 1.0$.
  The jet algorithm was applied in the laboratory frame.
A muon, identified in the central muon system, with transverse momentum
  $\ptmu > 2.5\,\GeV$ has to be found in a pseudorapidity range
  $-0.55 < \etamu < 1.1$. The muon has to be associated to one of the two
  highest energy jets.

To ensure a good muon reconstruction and suppress events with cosmic
muons, additional quality cuts were applied to the muon track;
in particular, the track must be associated to at least one hit
in the central silicon tracker to ensure a precise measurement of the
impact parameter $\delta$.

\section{Cross Section Definition and Measurement}

We have measured the cross section of the process $ep \to e b \bar b X \to ejj\mu X'$,
i.e. beauty production with the formation of two jets and 
the subsequent decay of a beauty hadron to a muon.
The muon may be produced by a direct semileptonic decay of a beauty hadron,
from a cascade decay, where the charm hadron decays semileptonically,
or from a $J/\psi$ or $\psi'$ decay.
The muon must be associated to either of the two highest $\pt$ jets.

The visible range is defined by
$Q^2< 1\,\GeV^2$,
$0.2 < y < 0.8$, 
$\ptmu > 2.5\,\GeV$,
$-0.55 <  \etamu < 1.1$
$\ptjets > 7 (6) \,\GeV$, and
$-2.5 < \etajets < 2.5$.   


To extract the beauty fraction, two quantities were used:
$\ptrel$, which is the transverse momentum of the muon with respect to the 
  axis of the most energetic jet, 
 reconstructed without the muon four vector,
 and
$\delta$, which is the impact parameter of the muon with respect to the primary
  vertex of the event.
The sign of $\delta$ is
determined in relation to the jet axis and defined such that it is 
positive for muons originating from a secondary vertex displaced along 
the jet direction.

For each bin, the two-dimensional distribution of these quantities
were fitted with three template distributions derived
from Monte Carlo simulations, taking into account the statistical 
uncertainties from
the data sample and the Monte Carlo templates \cite{Barlow:1993dm}; 
the templates were generated separately
for events containing only light ($u$, $d$, and $s$) quarks, charm
quarks, and beauty quarks, respectively.

The result of the fit is the relative amount of beauty induced events
$\fb$ in each analysis bin, from which the observed number of beauty events
$\Nb$ in the bin is calculated as
$
  \Nb = \fb \cdot \Nbin,
$
where $\Nbin$ is the total number of data events observed in the respective bin;
the data were then corrected for effects of detector resolution by a
matrix unfolding procedure, with a migration matrix determined from Monte Carlo
simulations.

\subsubsection*{Control Distributions}

\begin{figure}[bt]
\begin{center}
\setlength{\unitlength}{1cm}
\begin{picture}(12,4.4)
 \put(0,-0.5){\epsfig{file=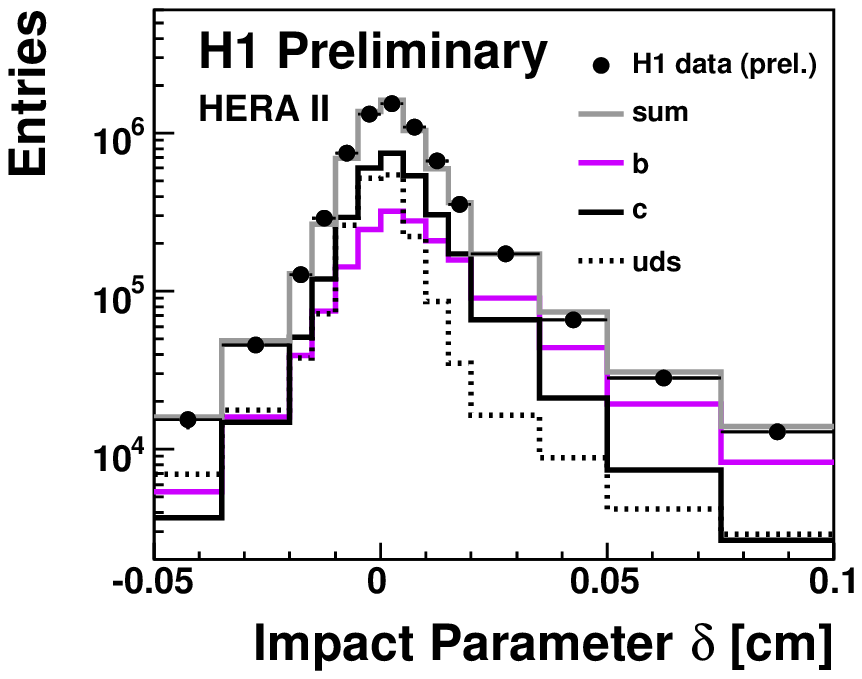,width=6cm}}
 \put(6,-0.5){\epsfig{file=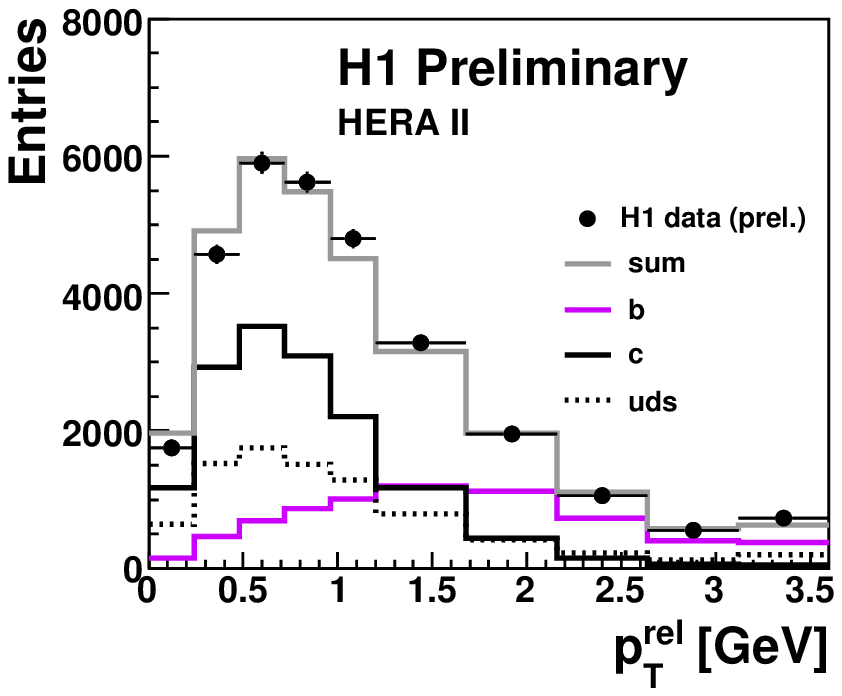,width=6cm}}
 \put(4.9,3.0){a)}
 \put(10.9,3.0){b)}
\end{picture}
\end{center}
    \caption{Distributions of a) the impact parameter
$\delta$ of the muon track and b) the transverse muon momentum
$\ptrel$ relative to the axis of the associated jet.
    Included in the figure are the estimated contributions of events
arising from beauty quarks (dark grey line), charm quarks (black line) and
light quarks (dotted line).}
 \label{fig:1}
\end{figure}

Fig.~\ref{fig:1} shows the distributions of the quantities used to
extract the beauty fraction from the data:
the impact parameter
$\delta$ of the muon track and the transverse muon momentum
$\ptrel$ relative to the axis of the associated jet.
Both quantities are described quite well by the Monte Carlo simulation.
In particular, the impact parameter distribution, which is very
sensitive to the detector resolution, is very well described, in the
region $\delta<0$ that is dominated by resolution effects as well as in
the region $\delta>0$, which shows the tails due to long-lived particles
from charm and beauty decays.

\subsubsection*{Systematic Uncertainties}

A number of sources of systematic errors were considered.
The overall normalization uncertainty comprises uncertainties on
the trigger efficiency, muon identification and track finding efficiencies,
and
the luminosity.
Additional uncertainties that affect the data differently in
various bins are:
The impact parameter resolution,
the reconstruction of the jet axis, 
the energy scale for hadrons of the calorimeter, 
the model uncertainties (estimated by using  CASCADE
  \cite{bib:cascade} instead of PYTHIA 6.2 \cite{bib:pythia}),
the uncertainty from the fragmentation process (estimated by using the Lund
 instead of the Peterson fragmentation function),
the uncertainty from the fragmentation fractions of $c$ and $b$ quarks into hadrons,
their branching ratios and lifetimes,
and the uncertainty on the modelling of $\pi$ and $K$ inflight
decays.
The resulting systematic uncertainty is $12\,\%$.

\section{Results}

\begin{figure}[p]
\setlength{\unitlength}{1cm}
\begin{center}
\begin{picture}(12,13.7)
 \put(0,0){\epsfig{file=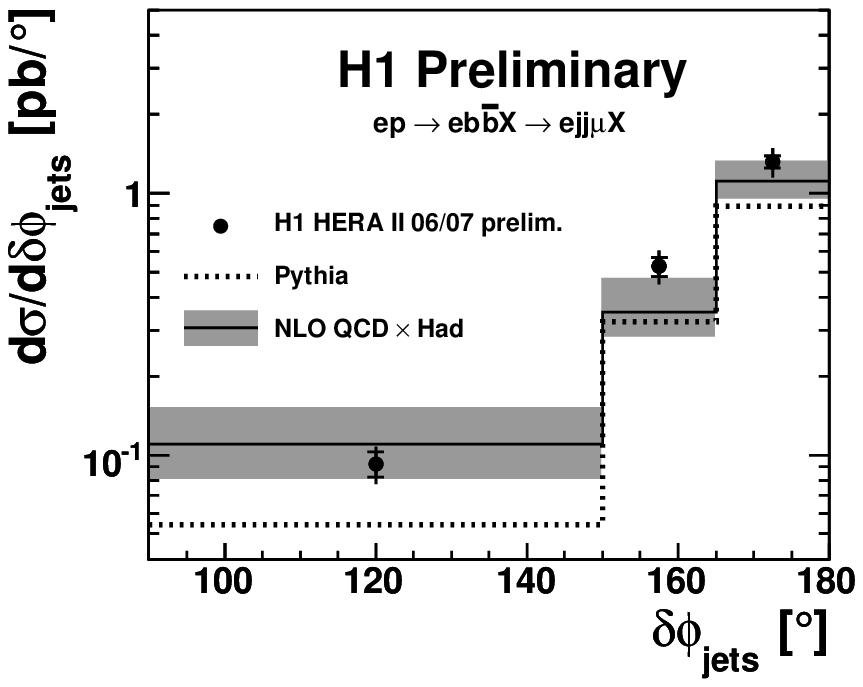,width=6cm}}
 \put(0,4.5){\epsfig{file=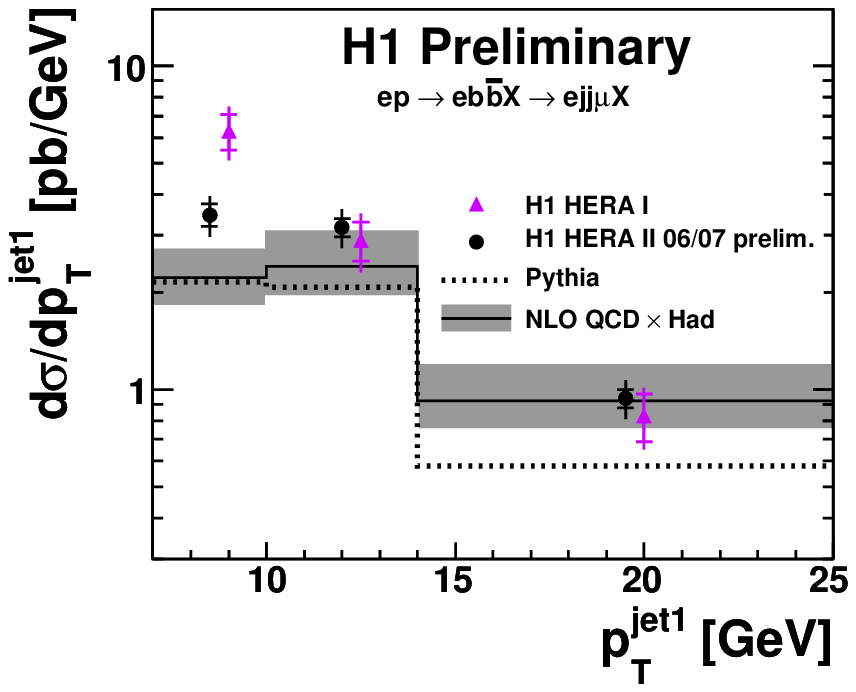,width=6cm}}
 \put(6,4.5){\epsfig{file=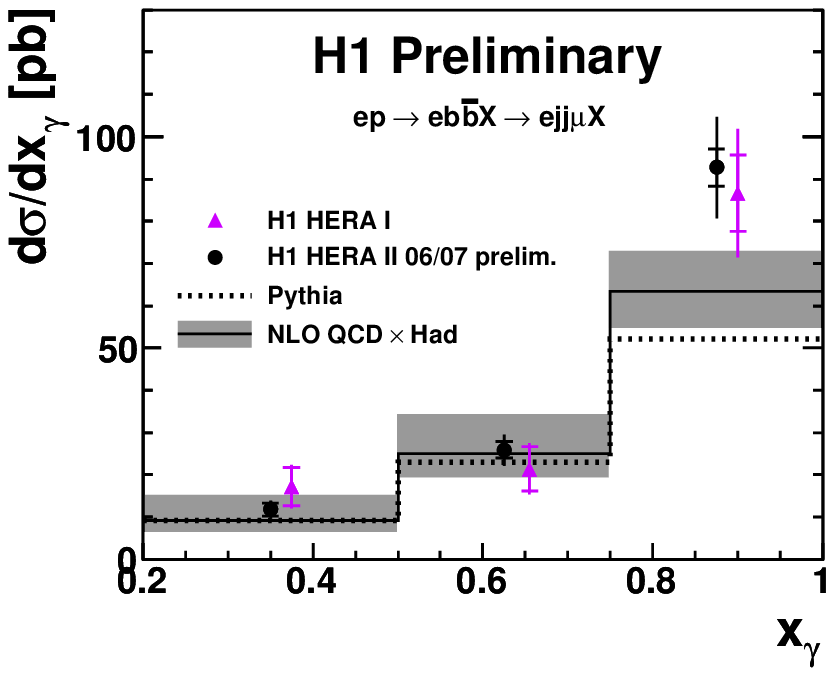,width=6cm}}
 \put(0,9){\epsfig{file=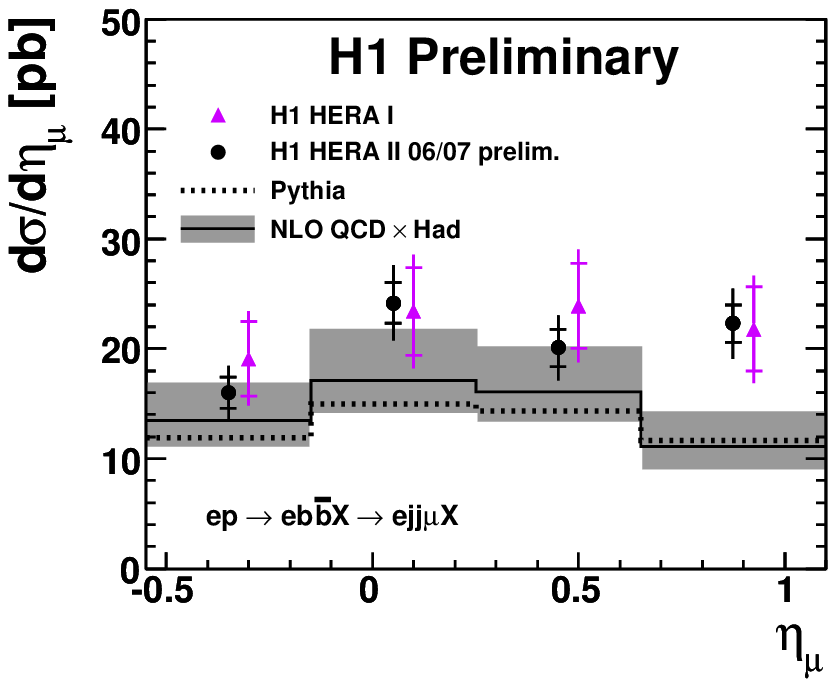,width=6cm}}
 \put(6,9){\epsfig{file=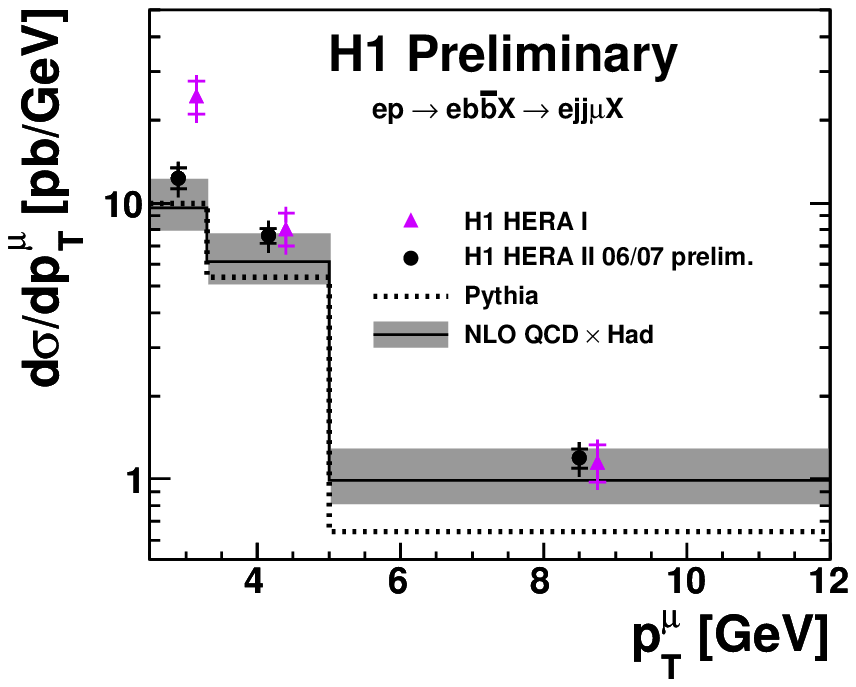,width=6cm}}
 \put(4.9,13){a)}
 \put(10.9,13){b)}
 \put(4.9,8.5){c)}
 \put(10.9,8.5){d)}
 \put(4.9,4){e)}
\end{picture}
\end{center}
    \caption{
Differential cross sections for the photoproduction process
$ep \to eb\bar b X \to ejj\mu X$ in the kinematic range 
$Q^2 < 1\,\GeV^2$, $0.2 < y < 0.8$, $\ptmu > 2.5\,\GeV$, 
$0.55 < \etamu < 1.1$, $\ptjets > 7(6)\,\GeV$ and $|\etajets| < 2.5$. The cross sections are
shown as functions of a) the muon pseudorapidity $\etamu$, b) the
muon transverse momentum $\ptmu$, c) the jet transverse momentum
$\ptjetone$ of the highest transverse momentum jet, d) the photon's momentum
fraction $\xgamma$ entering the hard interaction, 
and e) the azimuthal angle difference $\deltaphi$ between the jets.
 The inner error bars show the statistical error,
the outer error bars represent the statistical and systematic
uncertainties added in quadrature. The NLO QCD predictions
are corrected to the hadron
level (solid line) using the PYTHIA generator. The shaded
band around the hadron level prediction indicates the systematic
uncertainties as estimated from scale variations (see text).
Predictions from the Monte Carlo generator PYTHIA (dotted line) are also
shown.}
 \label{fig:2}
\end{figure}

The total cross section for the process $ep \to e b \bar b X \to ejj\mu X'$
in the visible range given above
has been measured to be
$$
  \sigma\sub{vis}\,(ep \to e b \bar b X \to ejj\mu X')  
  = 31.4 \pm 1.3 (stat.) \pm 3.8 (syst.) \,\pb.
$$
This result is somewhat lower than the published result \cite{Aktas:2005zc} from HERA-I,
which is \\
\mbox{%
$
  \sigma\sub{vis}
  = 38.4 \pm 3.4 (stat.) \pm 5.4 (syst.) \,\pb,
$}
but compatible within errors.
In comparison, the FMNR calculation yields
$
  \sigma\sub{vis}
  = 25.3^{+6.4}_{-4.7} \,\pb,
$
in agreement with the data,
and the PYTHIA prediction is
$
  \sigma\sub{vis}
  = 21.7 \,\pb.
$

Differential cross sections were also measured as a function
of the following quantities:
The transverse momentum of the muon $\ptmu$,
the transverse momentum $\ptjetone$ of the highest $\pt$ jet,
the pseudorapidity $\etamu$ of the muon,
\xgamma, the momentum fraction of the photon
entering the hard interaction, 
and $\deltaphi$, the difference in azimuthal angle between the two jets.

The differential cross sections also tend to be lower than the HERA-I
results, as shown in Fig.~\ref{fig:2}.
The largest discrepancies are observed for the differential measurements
in the lowest bins of $\ptmu$ and $\ptjetone$.
For these bins, the ratio between the measurements is 
about $2.5\,\sigma$ below unity if systematic uncertainties that do not cancel
between both measurements are taken into account. 
It has been checked that this discrepancy is not caused by differences in the analysis
method between the HERA-I and HERA-II analyses; we therefore attribute the difference to a
statistical fluctuation.

Fig.~\ref{fig:2}e) shows the difference in azimuthal angle, $\deltaphi$,
between the two jets. In leading order QCD the two outgoing quarks must be exactly
opposite in azimuthal angle, corresponding to $\deltaphi = 180^\circ$. 
Thus, values of $\deltaphi$ substantially lower than $180^\circ$ are indicative of the
presence of further final state gluons, and therefore this quantity is sensitive to
the description of gluon emission. PYTHIA, which employs parton showers to simulate 
the effect of multiple gluon emission, describes the shape of this observable
reasonably 
well, as does the fixed order calculation of FMNR, which allows at most one
hard gluon in the final state.

Overall, the data are reasonably well described  in shape by the predictions from PYTHIA,
but lie approximatly a factor $1.4$ above the PYTHIA prediction.
The NLO predictions, derived with the FMNR program, also lie systematically below the
data, but also describe the differential distributions well in shape.
In particular, the deficiency in the lowest bins 
of $\ptmu$ and $\ptjetone$ is not substantially larger than in the other bins,
in contrast to the findings of the HERA-I analysis. 
This observation agrees with the results from ZEUS \cite{bib:zeus-bmux}.

\section{Conclusions}

We have performed a measurement of the photoproduction of beauty quarks,
using events where at least one beauty hadron decays with a muon in the final state, and
two jets are visible in the detector,
in the phase space region defined by
$        Q^2< 1\,\GeV^2,
  0.2 < y < 0.8,
        \ptmu > 2.5\,\GeV,
  -0.55 < \etamu < 1.1,
        \ptjets > 7 (6) \,\GeV, $ and  
$   -2.5 < \etajets < 2.5.  $ 

The visible cross section has been measured to be
$
  \sigma\sub{vis}\,(ep \to e b \bar b X \to ejj\mu X')  
  = 31.4 \pm 1.3 (stat.) \pm 3.8 (syst.) \,\pb.
$
A NLO QCD calculation is in agreement with this measurement within the theoretical uncertainties.

Differential cross sections have been measured as function of the observables
$\etamu$, $\ptmu$, $\ptjetone$, $\xgamma$, and $\deltaphi$. 
The shape of these distributions is reasonably well described by 
the NLO QCD calculation as well as the PYTHIA LO Monte Carlo program.

At low values of $\ptmu$ and $\ptjetone$, the new measurement lies lower than 
the previous HERA-I measurement published by H1, and is thus better described by the NLO
predictions than the previous measurement.


\begin{footnotesize}


\end{footnotesize}


\end{document}